# Giant edge state splitting at atomically precise graphene zigzag edges

Shiyong Wang[1]*, Leopold Talirz[1]*, Carlo A. Pignedoli[1,2], Xinliang Feng[3], Klaus Müllen[3], Roman Fasel[1,4], Pascal Ruffieux[1]†

[1]*Empa, Swiss Federal Laboratories for Materials Science and Technology, 8600 Dübendorf, Switzerland*

[2]*NCCR MARVEL, Empa, Swiss Federal Laboratories for Materials Science and Technology, 8600 Dübendorf, Switzerland*

[3]*Max Planck Institute for Polymer Research, 55124 Mainz, Germany*

[4]*Department of Chemistry and Biochemistry, University of Bern, 3012 Bern, Switzerland*

*These authors contributed equally to this work.
†Corresponding Author: (PR) pascal.ruffieux@empa.ch

**Zigzag edges of graphene nanostructures host localized electronic states that are predicted to be spin-polarized. However, these edge states are highly susceptible to edge roughness and interaction with a supporting substrate, complicating the study of their intrinsic electronic and magnetic structure. Here, we focus on atomically precise graphene nanoribbons whose two short zigzag edges host exactly one localized electron each. Using the tip of a scanning tunneling microscope, the graphene nanoribbons are transferred from the metallic growth substrate onto insulating islands of NaCl in order to decouple their electronic structure from the metal. The absence of charge transfer and hybridization with the substrate is confirmed by scanning tunneling spectroscopy, which reveals a pair of occupied / unoccupied edge states. Their large energy splitting of $\Delta_{zz}$ = 1.9 eV is in accordance with *ab initio* many-body perturbation theory calculations and reflects the dominant role of electron-electron interactions in these localized states.**

Recent advances in the fabrication of precise graphene nanostructures open the door to the specific tailoring of the electronic properties of graphene-related materials. In the case of armchair graphene nanoribbons (AGNRs) it has been proven that the bottom-up fabrication allows controlling width and edge termination at the atomic level[1] and thus enables precise tailoring of the electronic band gap[2–4] and optical response[5] by a rational design of the molecular building blocks. Even more intriguing are the properties of graphene nanostructures involving zigzag edges for which different levels of theory predict the existence of spin-polarized edge states[6–10]. While a significant number of theoretical studies have investigated specific zigzag-related graphene nanostructures revealing spin filtering properties[11], half-metallic behavior[12] and spin confinement[13], the experimental results related to zigzag edged graphene nanostructures are extremely sparse and largely affected by limited structural precision and pronounced interaction with the substrate. Previous experimental studies of graphene zigzag edges have concentrated mainly on metal-adsorbed graphene nanostructures, where the low-energy edge states may interact with the nearby electron reservoir. For graphene nanoislands on Ir(111), edge states are found to be completely suppressed[14]. On less reactive surfaces, such as Au(111)[15–20], edge states have been observed for a variety of graphene nanostructures, even at interfaces between graphene and hexagonal boron-nitride[18,20]. However, the reported spectroscopic features of edge states vary greatly. For example, the energy splitting between occupied and empty edge states ranges from 0 eV[17] to 0.3 eV[19]. These values are much smaller than expected from electronic structure calculations for structurally perfect zigzag edges within many-body perturbation theory, which predict a splitting of 1.9 eV for the most strongly localized edge state[8]. Indeed, a recent study of the edges of graphene grown on silicon carbide[21] reports a substantial energy splitting of up to 1.2 eV, although the edges obtained from nanoparticle-assisted etching lack atomic precision. Reducing both, edge roughness and substrate interaction is thus a prerequisite for studying the intrinsic electronic and magnetic structure of graphene zigzag edges.

Here, we focus on the electronic properties of the atomically precise zigzag edges formed at the termini of bottom-up fabricated AGNRs. In order to achieve their electronic decoupling from the metal growth substrate we transfer the AGNRs onto NaCl islands by applying a scanning tunneling microscopy- (STM-) based multistep manipulation technique. Using scanning tunneling spectroscopy (STS), we find that electronic decoupling of the edge states establishes a large energy splitting between occupied and unoccupied edges states, which is in accordance with *ab initio* many-body perturbation theory calculations that we use to systematically assign zigzag-edge related edges states and the spatially and energetically separated nanoribbon bulk states.

## Results

We focus on short armchair GNRs of width $m=7$, which are synthesized with atomic-scale precision on a Au(111) single crystal surface using a recently established bottom-up method[1]. As sketched in Figure 1a, they are denoted as (7, $n$) GNRs, where $n$ specifies their length along the armchair direction. The (7, $n$) GNRs host two qualitatively different sets of electronic states. One set derives from the Bloch states of the bulk GNR, which are delocalized along the GNR. The short zigzag edges at the termini of the GNRs give rise to another set of states that are localized near the termini[10]. As sketched in Figure 1b-c, these edge states (also called Tamm states[22]) are energetically isolated from the delocalized bulk states of the GNR, thus offering an experimental advantage over graphene nanostructures with long zigzag edges, where the energies of edge-localized and delocalized states are predicted to overlap[6,8]. Previous STS investigations of (7, $n$) GNRs on Au(111) indicated only one, possibly degenerate, edge state near the Fermi level, which has been explained in terms of hole-doping of the GNR[17]. To access their intrinsic properties, the GNRs thus need to be transferred onto insulating substrates, which is also required for future GNR-based applications[23]. In order to characterize the intrinsic electronic structure of (7, $n$) GNRs, the transfer onto an insulating substrate needs to occur after the synthesis, which relies on the catalytic activity of the metal surface. Here, we use atomically thin insulating NaCl films on metals which, in contrast to bulk insulating substrates, enable the investigation of the electronic properties of the GNRs by STM/STS while considerably lowering their interaction with the metal substrate[24]. Through a novel four-step STM manipulation routine (see methods and supporting information), we transfer ribbons with lengths ranging from 2 to 10 nm onto a monolayer of NaCl without introducing any defects. The transfer process relies on the weak adhesion of defect-free (7, $n$) GNRs to the Au(111) growth substrate, which allows for lateral manipulation and controlled pick-up of individual GNRs by the STM tip[22]. A typical STM scan of a (7, 20) GNR on NaCl is shown in Fig. 2b. When positioning the STM tip above a zigzag end of the decoupled (7, 20) GNR, the differential conductance ($dI/dV$) spectrum shows two peaks centered at – 0.5 V and 1.3 V (cf. figure 2c). The peaks are well separated from the Fermi energy, excluding the possibility of (partial) charge transfer. STM images taken at these bias voltages clearly associate the peaks with electronic states localized at the zigzag terminated ends. As shown in figure 2d, the shapes of filled and empty edge states are essentially identical. Their characteristic features, such as a broadening towards the very end of the GNR as well as the protrusions at the outermost carbon atoms, are in excellent agreement with the orbital den-sities of the corresponding states in Kohn-Sham DFT calculations for freestanding GNRs (cf. figure 2e).

Being a theory of the electronic ground state, Kohn-Sham DFT, however, is not designed to describe the charged excitations that take place in STS, which involve the addition/removal of electrons to/from the sample. While the orbitals of the non-interacting Kohn-Sham system are often found to be accurate approximations of the corresponding quasiparticle wave functions[25], the Kohn-Sham orbital energies are known to deviate significantly from quasiparticle excitation energies in many bulk insulators and molecules. In particular, the Kohn-Sham gap of standard semi-local DFT functionals (and even of the exact functional[26]) can severely underestimate the fundamental gap, defined as the difference between the ionization potential and the electron affinity. An accurate description of the fundamental gap needs to properly account for the interaction of the additional charge with the remaining electrons. This many-body effect of dynamical screening is captured naturally by many-body perturbation theory in the *GW* approximation. The framework provides accurate fundamental gaps, both for of bulk insulators[27] and molecules[28], and has been applied successfully to GNRs of infinite length[8].

Here we perform *ab initio GW* calculations for finite (7,*n*) GNRs, which can be viewed as open-shell molecules with one, singly occupied state localized at each terminus. Their fundamental gap coincides with the energetic splitting $\Delta_{ZZ}$ between the occupied and empty edge-localized states. A significant Kohn-Sham gap opens only in the spin-unrestricted formalism, where breaking of spin symmetry gives rise to staggered sublattice potentials[7]. Using the semi-local PBE functional[29] a Kohn-Sham gap of $\Delta_{ZZ}^{KS}$ = 0.54 eV is obtained for lengths $n \geq 12$. Starting from the PBE orbitals and orbital energies, we compute quasiparticle corrections in the $G_0W_0$ approximation. The fundamental gap $\Delta_{ZZ}$ is found to converge rapidly as a function of length, yielding a value of $\Delta_{ZZ}^{GW}$ = 2.8 ± 0.1 eV for $n \geq 12$ (see Figure 3g), exceeding the Kohn-Sham gap by more than a factor of five. Note also that $\Delta_{ZZ}^{GW}$ is larger than the $G_0W_0$ gap of max. 1.9 eV between states localized at *extended* zigzag edges[8], as expected from the additional confinement along the zigzag direction (for the link between the edge state of the (7,*n*) GNR and the edge states of extended zigzag edges, please see the supporting information). Direct comparison to experiment would require the inclusion of dynamical screening not only by the electrons of the GNR itself, but also by those of the NaCl monolayer and the underlying Au substrate, which is expected to lead to significant reduction of the fundamental gap[2]. Due to computational constraints, we do not describe screening by the substrate quantitatively here (see [2,30] for studies modeling such ef-fects), but point out that the experimentally observed gap of $\Delta_{ZZ}$ = 1.9 eV is fully compatible with a fundamental gap of 2.8 eV, reduced by screening from substrate electrons.

In order to investigate finite size effects, ribbons of different lengths have been moved onto NaCl islands and inspected. Figure 3a to 3d show STM topographies (upper panel) and STS maps (lower panel) of filled and empty edge states of (7, 12), (7, 16), (7, 20) and (7, 48) GNRs, respectively. In accordance with measurements on Au(111) as well as theory[17,31], the edge states are found to be localized near the zigzag termini with a typical extent of 1.5 nm (see supporting information for a tight binding analysis). Over the length range of 3 - 10 nm investigated here, $\Delta_{ZZ}$ is essentially independent of the separation between the zigzag edges, in accordance with the $G_0W_0$ predictions (cf. Figure 3e). Similar observations have also been made on long, chemically etched zigzag GNRs on SiC, where a constant gap of 0.12 eV is reported for ribbons wider than 3 nm[21]. In the case of long zigzag GNRs, however, the edge-localized states overlap energetically with delocalized bulk states[8], making it difficult to distinguish between the two in STS. For (7,$n$) GNRs, the additional quantum confinement at the short zigzag edges selects one particular wavelength along the zigzag edge and the corresponding edge state is energetically isolated from the delocalized states of (7, ∞) GNRs.

Next, we discuss the delocalized bulk electronic states of decoupled (7, $n$) GNRs. Their width of 7=3x2+1 carbon dimer rows identifies (7, $n$) GNRs as a member of the 3$m$+1 family of armchair graphene nanoribbons, which has the largest band gaps[7,8]. As shown in Figure 3f, $dI/dV$ spectra taken at the center of a (7, 12) GNR show two sharp peaks at –1.2 V and 2.3 V. These peaks indicate the onsets of the highest occupied and lowest unoccupied bulk states, yielding a bulk band gap $\Delta_{ac}$ = 3.5 eV. In contrast to the energy splitting $\Delta_{ZZ}$ of the localized edge states, we find that the bulk band gap $\Delta_{ac}$ decreases continuously from 3.5 eV for the shortest (7, 12) GNR to 2.9 eV for the longest (7, 48) GNR under study, as shown in Figure 3g. This trend is rationalized by the decreasing longitudinal confinement of the associated bulk states that extend throughout the GNR. As suggested by Figure 3g, the value of 2.9 eV measured for a GNR of 10 nm length is converged within experimental accuracy, in agreement with length-dependent band gap studies of (7,n) GNRs on Au(111)[32]. Many-body perturbation theory calculations in the GW approximation predict a band gap of 3.7 ± 0.1 eV for freestanding (7, ∞) GNRs[2,8]. While the observed band gap of $\Delta_{AC}$ = 2.9 eV is still significantly below the value predicted for the free-standing (7, ∞) GNR, it is significantly larger than the 2.4 eV measured for the (7, ∞) GNR on Au(111)[32], thus indicating considerably reduced screening by substrate electrons.

The dispersion of the electronic states of a decoupled (7,48)-GNR has been determined via Fourier transformed (FT) STS[32]. Figure 4a shows a grid of STS spectra taken along one armchair edge. In the color map, both the edge states and the bulk states can be resolved. The bulk states show standing waves arising from scattering at the termini of the GNRs, in good agreement with the corresponding DFT-based FT-STS simulation of a (7,48)-GNR (Figure 4b). For example, at -1.4 V bias, four nodes are observed along both armchair edges. With decreasing bias, we observe three nodes, two nodes and one node at -1.2, -1.1 and -1.0 V, respectively (see supplement for constant-current STS maps). In order to quantitatively investigate the electronic band dispersion, we perform a discrete Fourier transform to reciprocal space as shown in Figure 4b. One occupied band and two unoccupied bands can be resolved, with effective masses of 0.32 ± 0.04, 0.35 ± 0.10, and 0.20 ± 0.05 $m_e$ respectively ($m_e$ is the free electron mass). We note that the bands appear with different intensity in STS due to the finite tip-sample distance and refer interested readers to the corresponding study on Au(111) for details[32]. While the effective masses are slightly smaller than those determined for the (7, ∞) GNRs directly on Au(111)[32], the respective error bars overlap, indicating that the effective masses are largely unaffected by the electronic decoupling, despite the accompanying strong increase of the band gap. Figure 4c shows the DFT band structure of the (7, ∞) GNR and the corresponding many-body corrections within the $G_0W_0$ approximation. While the band gap is found to open from 1.6 eV to 3.7 eV[8], the effective masses are found to decrease only by ~10%. This finding is consistent with earlier work on graphene where many-body corrections to the local-density approximation give rise to a small increase of the dispersion near the Fermi energy[33].

## Conclusion

We have demonstrated the transfer of atomically precise (7,n) graphene nanoribbons from Au(111) to an insulating monolayer of NaCl without introducing any defects. The delocalized electronic states of (7,n) GNRs are separated energetically from the states localized at the zigzag termini, making it possible to investigate the effect of electronic decoupling on both classes of states separately. Using scanning tunneling spectroscopy, the band gap between delocalized states is found to increase from 2.4 eV to 2.9 eV upon electronic decoupling, while no significant modification of the effective masses is observed. Furthermore, we find an energetic splitting of the states localized at the zigzag termini that has long been predicted for freestanding zigzag edges, but is missing completely in (7,n) GNRs on Au(111). Both its substantial size of $\Delta_{ZZ}$ = 1.9 eV and its independence on edge separation down to

3 nm are in agreement with *ab initio* many-body perturbation theory calculations of the GNRs' intrinsic electronic structure, indicating that decoupling by a single layer of NaCl indeed allows to study the intrinsic properties of graphene zigzag edges. We therefore expect that the experimental strategy established here will particularly benefit the eagerly awaited exploration of the low-energy spin physics at graphene zigzag edges using spin-sensitive methods.

**Methods**

Sample preparation and STM measurements were performed in an ultrahigh vacuum system (base pressure $1\times10^{-10}$ mbar) using an Omicron low-temperature STM. The Au(111) substrate was cleaned by standard argon sputtering and annealing cycles. The GNRs were grown on Au(111) following the recipe by Cai et al[1]. The sample temperature for the cyclodehydrogenation step was chosen such as to yield monohydrogenated termini[31]. NaCl powder was thermally evaporated at the sample held at room temperature. Immediately after NaCl deposition, the sample was transferred to the STM chamber and cooled down to 5K for characterization. This results mostly in NaCl monolayers, as judged by their apparent height of 2.2 Å. In order to transfer a GNR onto NaCl, the GNR is picked up at one end by the STM tip[22]. Together with the GNR, the tip is then moved laterally above the NaCl monolayer, while the other end of the GNR still remains physisorbed on Au(111). After applying a voltage pulse of 3.0 V to release the ribbon, the tip is used to push the GNR fully onto the NaCl monolayer (see supporting information). Transfer of GNRs onto bilayer NaCl was found to be more challenging due to lack of stable adsorption of the GNRs partially on NaCl bilayers. The *dI/dV* spectra were recorded using the lock-in technique ($U_{rms}$ = 20 mV).

Electronic structure calculations within the framework of density functional theory were performed with the PBE exchange-correlation functional[29]. Band structure calculations were carried out using the Quantum ESPRESSO package[34] (details in the supporting information). The electronic structure of the (7,48) GNR was calculated with the CP2K code[35], which expands the electronic wave functions on an atom-centered Gaussian-type basis set. After extrapolating the Kohn-Sham orbitals into the vacuum region[36], STS simulations were performed in the Tersoff-Hamann approximation[37] on a plane parallel to the planar GNR. Quasiparticle corrections were computed in the $G_0W_0$ approximation using the BerkeleyGW package[27,38]. The static dielectric matrix was calculated using a rectangular Coulomb-cutoff along the aperiodic directions[39] and extended to finite frequencies via the generalized plasmon pole model[27]. In the calculation of the self-energy, the static remainder approach was used to speed up the convergence with respect to the number of empty bands[40] (more details in the supporting information).

**Acknowledgements**

This work was supported by the Swiss National Science Foundation, by the Office of Naval Research BRC Program and by a grant from the Swiss National Supercomputing Centre (CSCS) under project ID s507.

**Author Contributions**

S.W., R.F and P.R conceived the experiments. The Mainz group synthesized the molecular precursors. S.W. performed the scanning probe experiments. L.T. and C.A.P. performed the simulations.

S.W., L.T., and P.R wrote the paper. All authors discussed the results and implications and commented on the manuscript at all stages.

**Competing financial interests**

The authors declare no competing financial interests.

**Figure Captions**

**Figure 1 | Electronic structure of finite GNRs**. **a**, Cutting graphene into nanoribbons with different edge topologies. Indices ($m$, $n$) are used to denote the dimensions of a nanoribbon along the zigzag ($m$) and armchair direction ($n$), respectively. **b**, Sketch of energy levels for a finite (7, 12) GNR, with $\Delta_{ac}$ and $\Delta_{zz}$ indicating the bulk band gap and the splitting of the localized states at the zigzag edges, respectively. **c**, Kohn-Sham spin-orbitals of edge-localized states and energetically closest bulk states. Electrons with different spins are localized at opposing zigzag edges.

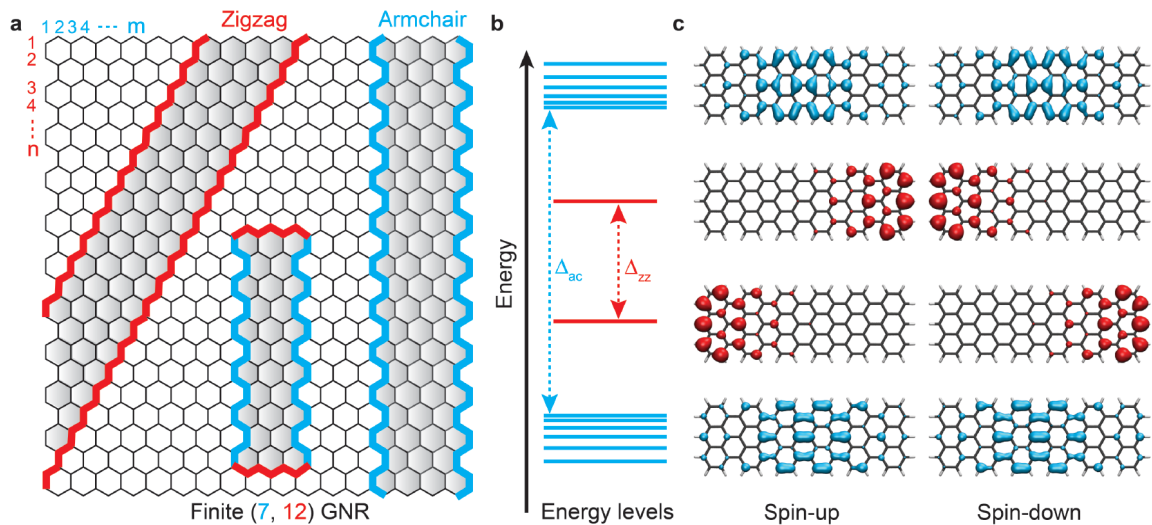

**Figure 2 | Electronic structure of (7, 20) GNR on NaCl monolayer. a,** Structural model of a (7, 20) GNR. **b,** STM topography image of a (7, 20) GNR transferred onto a NaCl monolayer island through STM manipulation ($U$ = -1.0 V, I = 30 pA). **c,** Differential conductance spectra measured in the center (blue) and at a zigzag end (red) of the decoupled (7, 20) GNR. Inset: STM topography image at sample bias in the band gap of the ribbon ($U$ = - 0.1 V, I = 30 pA). **d,** STM topography images showing the orbital shapes of the occupied edge state (left, $U$ = - 1.0 V, I = 30 pA) and the unoccupied edge state (right, $U_e$ = 1.4 V, I = 30 pA). **e,** Local density of states of corresponding Kohn-Sham orbitals at 4 Å distance above the GNR.

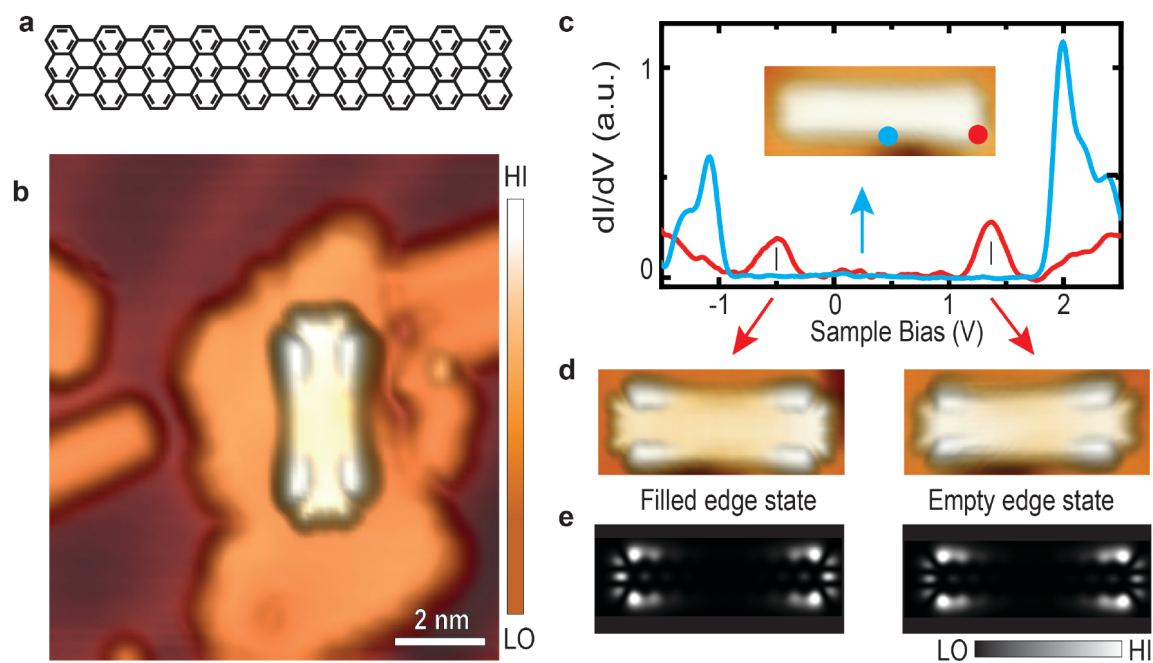

**Figure 3 | Length-dependent electronic structure of decoupled (7, *n*) GNRs. a - d,** STM topography images and STS maps of empty and filled edge states of decoupled (7, 12), (7, 16), (7, 20) and (7, 48) GNRs on NaCl monolayer islands (*U* = -0.1 V, I = 30 pA). **e - f,** Differential conductance spectra taken at the terminus (**e**) and at the center (**f**) of each ribbon shown in **a - d**. **g,** Bulk band gap $\Delta_{ac}$, edge state splitting $\Delta_{zz}$, and calculated GW splitting $\Delta_{zz}$(GW) as a function of inverse GNR length (the dashed line serves as a guide to the eye).

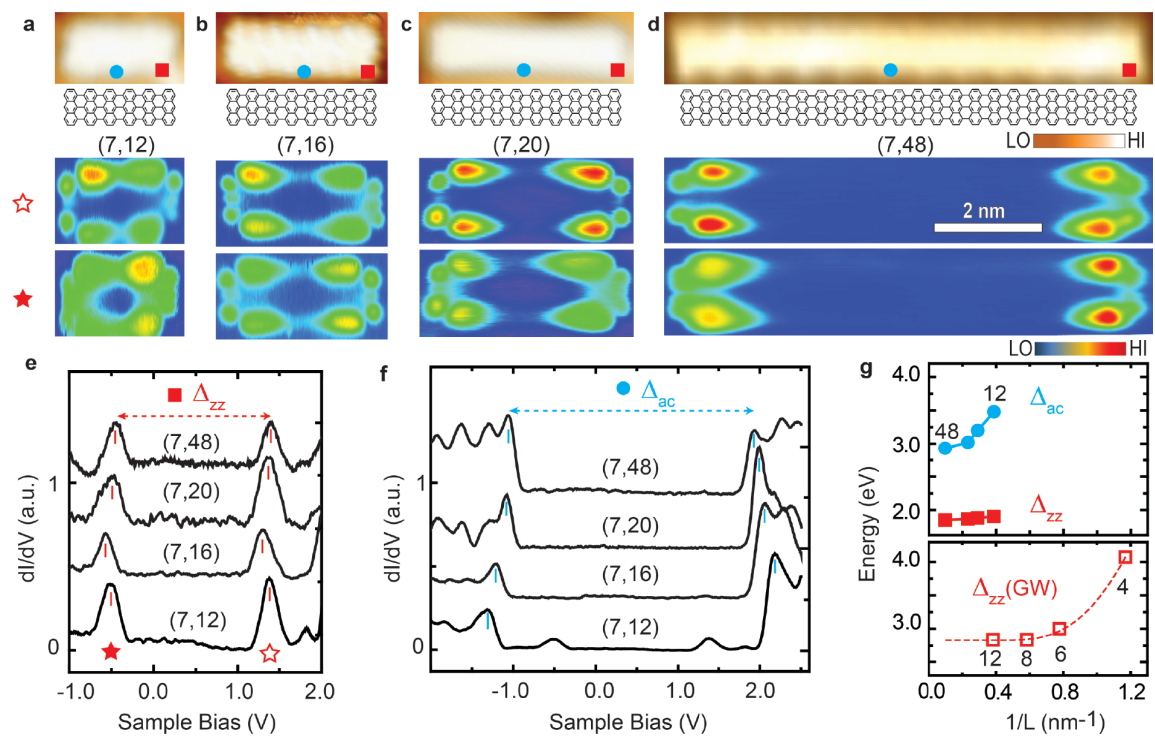

**Figure 4 | Band structure of decoupled (7, 48) GNR. a,** Left panel: Grid of dI/dV spectra (spaced 0.15 nm) taken along the armchair edge of a decoupled (7, 48) GNR. Right panel: Fourier transformed map revealing one occupied band and two unoccupied bands near the Fermi level. **b,** STM topography of a decoupled (7,48) GNR ($U$ = -1 V, I= 30 pA). **c,** Left panel: DFT-based LDOS of (7, 48) GNR at 4 Å tip-sample distance (integrated across the ribbon). Right panel: Fourier transformed LDOS with DFT bands of inifinite ribbons superposed as dashed red lines. **d,** DFT and GW band structure of (7, ∞) GNR, aligned at the center of the gap (zero energy).

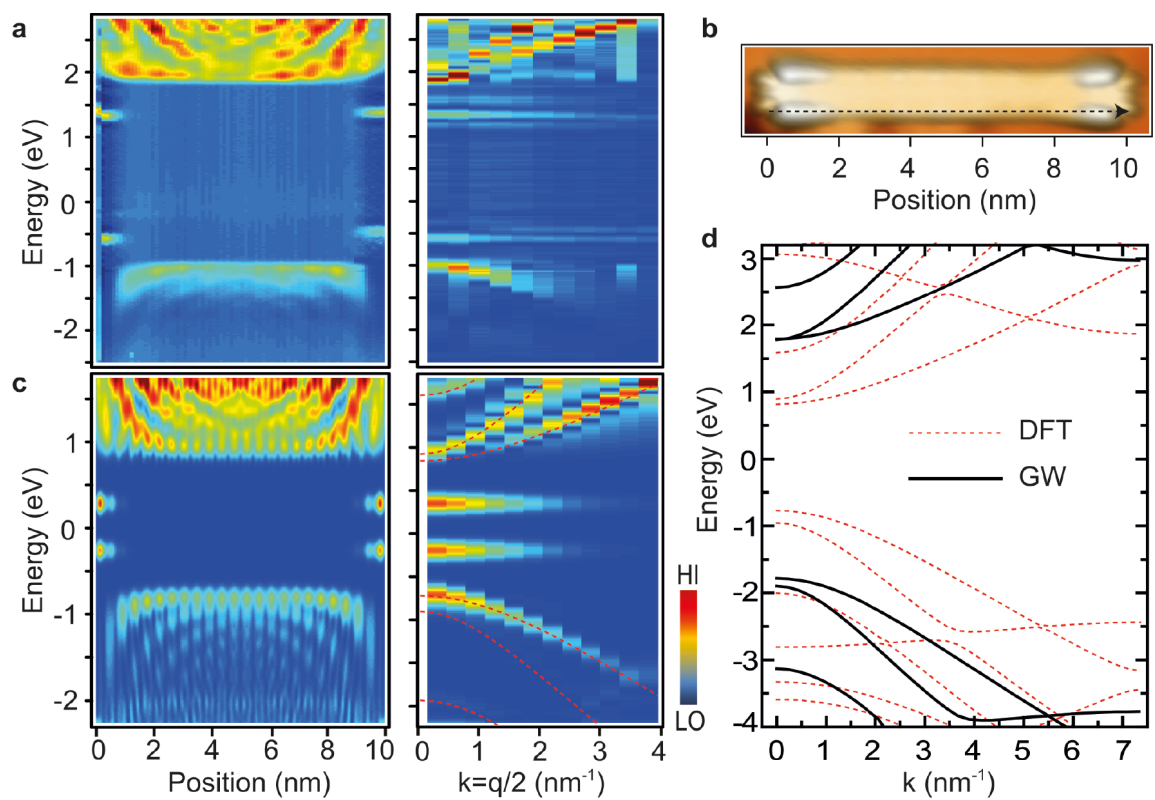

# Giant edge state splitting at atomically precise zigzag edges

Shiyong Wang[1]*, Leopold Talirz[1]*, Carlo A. Pignedoli[1,2], Xinliang Feng[3], Klaus Müllen[3], Roman Fasel[1,4], Pascal Ruffieux[1†]

[1]*Empa, Swiss Federal Laboratories for Materials Science and Technology, 8600 Dübendorf, Switzerland*

[2]*NCCR MARVEL, Empa, Swiss Federal Laboratories for Materials Science and Technology, 8600 Dübendorf, Switzerland*

[3]*Max Planck Institute for Polymer Research, 55124 Mainz, Germany*

[4]*Department of Chemistry and Biochemistry, University of Bern, 3012 Bern, Switzerland*

## Supplementary Information

S1. Transfer of GNRs onto NaCl monolayer islands

S2. Particle-in-a-box states

S3. Electronic structure of decoupled defective GNR

S4. Computational details

S5. Edge states within tight binding

References

**S1. Transfer of GNRs onto NaCl monolayer islands**

Individual (7,n) GNRs are physisorbed on Au(111), where they can be moved laterally by an STM tip. However, direct pushing of ribbons onto NaCl monolayer islands was found impossible. The apparent height of NaCl (2.2 Å) is higher than that of GNRs (1.8 Å) and, moreover, the NaCl islands themselves are physisorbed on Au(111) and easily moved by an STM tip. Recently, Koch *et al.* have demonstrated that individual ribbons can be picked up by an STM tip[1]. Inspired by their work, we have developed an STM manipulation routine (cf. Figure S1) to transfer bottom-up fabricated GNRs from the metal substrate onto insulating NaCl monolayers in order to access their decoupled electronic structure. We have successfully achieved the transfer of GNRs with lengths ranging from 3 nm to 10 nm without introducing any defects. The STM manipulation routine consists of four steps: (1) pick-up of one end of a GNR by approaching and retracting the STM tip with low bias (~ -50 mV); (2) lateral movement of the tip (together with the GNR) above the NaCl island; (3) release of the ribbon by a 3.0 V voltage pulse, leaving the GNR partially adsorbed on NaCl and partially on the metal surface; and (4) lateral positioning of the GNR in order to have it entirely adsorbed on the NaCl island. Step 3 was introduced because releasing a GNR from the tip was found much easier when one end of the GNR is still adsorbed on Au(111). Transfer of GNRs onto bilayer NaCl has also been attempted, but was found to be much more challenging due to a lack of stable adsorption of the GNRs on NaCl bilayers in step 3.

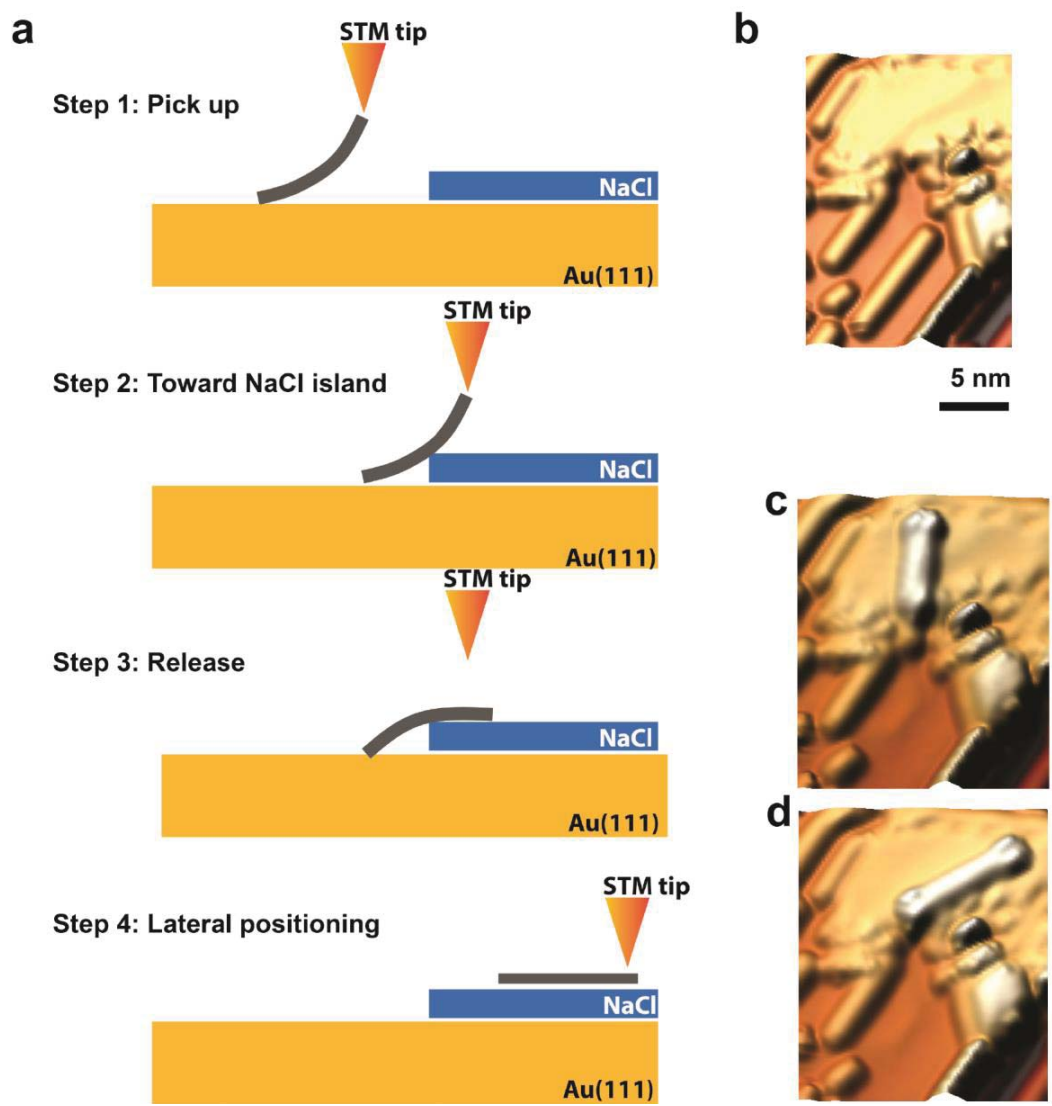

**Figure S1 | STM manipulation routine for transfer of GNRs onto NaCl monolayers. a,** Schematic illustration of STM manipulation. **b,** STM image ($U_{sample}$ = -0.1 V, I = 30 pA) of a (7, 48) GNR before manipulation. **c,** STM image ($U_{sample}$ = -0.1 V, I = 30 pA) after step 3 showing the ribbon partially adsorbed on NaCl. **d,** STM image ($U_{sample}$ = -0.1 V, I = 30 pA) after step 4 showing the ribbon fully adsorbed on the NaCl monolayer.

## S2. Particle-in-a-box states

The constant-current STS maps in Figure S2 show a local density of states (LDOS) distribution that can be understood with a simple quantum well picture. For example, at -1.4 V bias, four nodes are observed along both armchair edges. With decreasing bias, we observe three nodes, two nodes and one node at -1.2, -1.1 and -1.0 V, respectively. Similar features are observed for the unoccupied states, where the situation is more complex due to the overlap of two bands with different STS intensity. We note that the localization of the STS intensity near the armchair edges of the GNR, as seen in Figure S2 c-d, does not reflect an edge-localized nature of the underlying electronic states of the GNR. Due to the significant tip-sample distance of several Angstroms, the STS maps are defined by the exponential tails of the GNR orbitals, whose shape may differ significantly from that of the orbital close to the GNR. These effects are included in the STS simulations presented in Figures 2e and 4c of the main text and are described in more detail in reference[3].

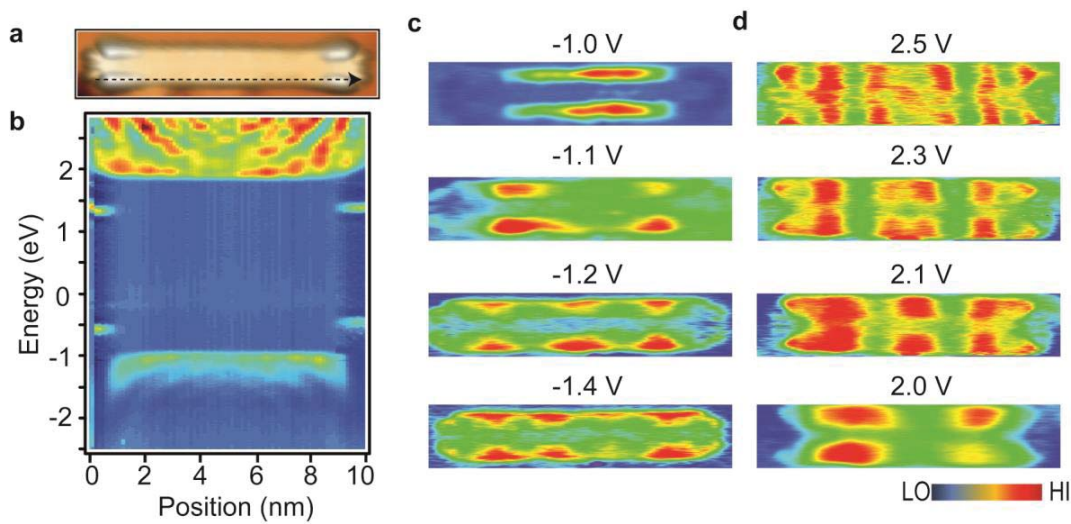

**Figure S2 | Particle-in-a-box state of a (7, 48) GNR. a,** STM image ($U_{sample}$ = -1 V, I = 30 pA) of a (7, 48) GNR. **b,** Grid of dI/dV spectra (spaced 0.15 nm) along the decoupled (7,48) GNR. **c,** constant-current dI/dV maps at different negative bias voltages exhibiting particle-in-a-box features due to quantum confinement. **d,** constant-current dI/dV maps at different positive bias voltages.

**S3. Electronic structure of decoupled defective GNR**

Besides a majority of defect-free (7, n) GNRs, the on-surface synthesis yielded also some defective ribbons. Figure S3 a shows a typical type of defect. The GNR consists of six bianthryl units, the last of which is slightly tilted with respect to the GNR axis. The atomic structure of this type of defect is known from previous non-contact atomic force microscopy measurements[2] and a corresponding structural model is superimposed onto the STM image of Figure S3 a. Edge states are found to be present both at the pristine left and the tilted right end (cf. Figure S3 c). Figure S3 d shows dI/dV spectra taken at different positions of the ribbon. As expected for a GNR of this length, an energy splitting of $\Delta_{zz} = 1.87$ eV $\approx$ 1.9 eV is observed at the pristine end, whereas a significantly larger splitting of $\Delta_{zz} = 2.18$ eV is observed at the tilted end. Interestingly, the energy levels of the occupied edge states are identical for both ends, while the energy level of the empty edge state shifts up significantly at the tilted end. Tilting of the terminal bianthryl unit may increase quantum confinement, both for the edge state on the terminal bianthryl unit and for the bulk states of the untilted part of the GNR, resulting in an increase of the energy splitting $\Delta_{zz}$ as well as of the bulk band gap. When positioning the STM tip over the central part of the ribbon, the dI/dV spectrum yields a bulk band gap of 3.05 eV, similar to the band gap of (3.02 eV) determined for a defect-free GNR with five bianthryl units.

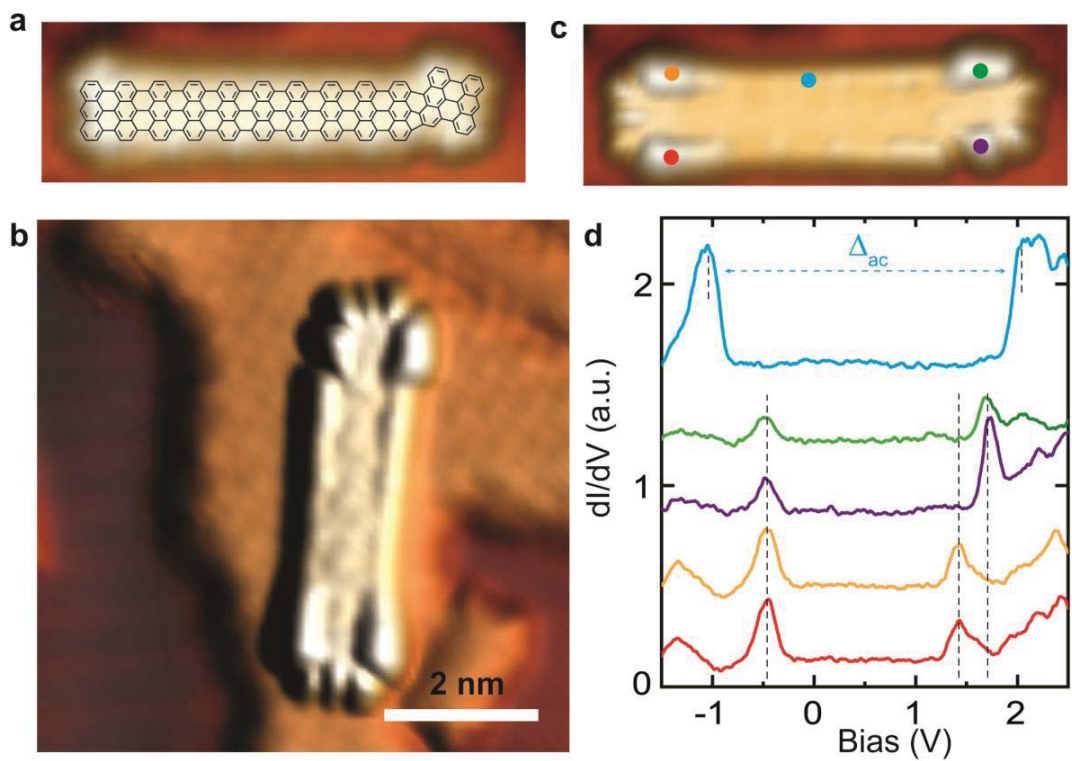

**Figure S3 | Electronic structure of a defective ribbon. a,** STM image ($U_{sample}$ = 0.5 V, I = 30 pA) of a defective GNR consisting of 12 anthryl units superimposed by a structure model. **b,** STM image ($U_{sample}$ = -1 V, I = 30 pA) showing the ribbon fully on NaCl monolayer island. **c,** STM image ($U_{sample}$ = -1 V, I = 30 pA) showing the shape of edge states. **d,** Differential conductance spectra measured at different positions as marked by the colored dots in **c**.

## S4. Computational details

Electronic structure calculations were performed within the framework of density functional theory (DFT), using the PBE generalized-gradient approximation to the exchange-correlation functional as implemented in the Quantum ESPRESSO package[4]. The Kohn-Sham orbitals were expanded on a plane-wave basis set with energy cutoff at 150 Ry, using norm-conserving pseudopotentials.

For the finite (7,n) GNRs, the simulation cell was chosen at least twice as large as the $10^{-5}/a_0^3$ isosurface of the charge density in order to enable the Coulomb-cutoff technique[7] in the GW calculations. For the (7, ∞) GNR, 18 Å of vacuum were introduced in the directions perpendicular to the GNR axis, while 16 k-points were used to sample the first Brillouin zone. The lattice parameter of the (7,∞) GNR was determined to be 4.285 Å and atomic positions were relaxed until the forces acting on the atoms were below 3 meV/Å.

Quasi-particle corrections were computed within the framework of many-body perturbation theory, using the $G_0W_0$ approximation to the self-energy as implemented in the BerkeleyGW package[5,6]. The electronic structure from DFT was recalculated using 60 Ry plane-wave cutoff (and 64 k-points in the first Brillouin zone for the (7, ∞) GNR). We have computed sufficient numbers of empty states to cover the energy range up to 2.1 Ry (infinite GNR) and 1.6 Ry (finite GNRs) above the highest occupied band. The static dielectric matrix $\varepsilon$ was calculated in the random phase approximation with 8 Ry cutoff for the plane-wave basis. $\varepsilon^{-1}$ was extended to the real frequency axis using the generalized plasmon-pole model by Hybertsen and Louie[5]. A rectangular Coulomb-cutoff was employed along the aperiodic dimensions as described in reference [7]. In the calculation of the self-energy, the static remainder approach was used to speed up the convergence with respect to the number of empty bands[8].

Table S1 summarizes the values for the Kohn-Sham gap and the $G_0W_0$ gap of finite (7,n) GNRs. Note that using the PBE functional, a ground state with spin-polarized edge states is found only from length

n=6 onwards, while no spin polarization is found for the ultrashort GNRs of lengths n=2 (anthracene) and n=4 (bisanthene). This is in agreement with previous results obtained by the complete active space self-consistent field (CASSCF) method[9].

|  | n=2 | n=4 | n=6 | n=8 | n=12 |
| --- | --- | --- | --- | --- | --- |
| $\Delta_{zz}$ (PBE) [eV] | 2.33 | 0.90 | 0.58 | 0.55 | 0.54 |
| $\Delta_{zz}$ (G$_0$W$_0$) [eV] | 6.49 | 4.03 | 3.00 | 2.86 | 2.85 |

**Table S1 | Edge state splitting of finite (7,n) GNRs.** Comparing $\Delta_{zz}$ obtained from the PBE Kohn-Sham gap and from the G$_0$W$_0$ gap. GNR length is specified by the number of zigzag lines n, as defined in Figure 1a of the main text.

### S5. Edge states within tight binding

Single-orbital nearest-neighbor tight binding is a simple, yet instructive model that provides qualitative insights into graphene's π-electronic structure[10]. The corresponding Hamiltonian is given by

$$\mathcal{H} = -t \sum_{<ij>} c_j^\dagger c_i + c_i^\dagger c_j$$

where the sum extends over distinct pairs of nearest neighbors, $c_j^\dagger, c_i$ are the electron creation and annihilation operators on carbon sites $i, j$, and $t \approx 3$ eV is the hopping matrix element between nearest neighbors.

This model can be solved analytically both for graphene and for armchair and zigzag graphene nanoribbons[11]. Figure S4a shows the resulting band structure of graphene, projected onto the zigzag direction. The solid black line indicates the non-bonding, singly occupied zero-energy states localized at the zigzag edge of a semi-infinite graphene sheet, as sketched in Figure S4b. The degree of localization of these states depends on the wave vector k along the zigzag direction. Perpendicular to the zigzag edge, the wave functions $\psi_k$ decay exponentially with decay constant $\alpha_k = -2 \ln \left( 2 \cos \left( \frac{k a_{zz}}{2} \right) \right) / a_{ac}$,[11] ranging from complete delocalization at $k = \frac{\pi}{a_{zz}} \frac{2}{3}$ ($\alpha_k = 0$) to complete localization at $k = \frac{\pi}{a_{zz}}$ ($\alpha_k = \infty$). Here, $a_{ac} = \sqrt{3} a_{zz} = 3a$, where $a \approx 0.142$ nm denotes the carbon-carbon bond length in graphene.

Figure S4b shows a superposition of two such Bloch states with wavelength $\lambda = 8a_{zz}$ along the zigzag edge, corresponding to crystal wave vectors $k_0^{\pm} = \frac{\pi}{a_{zz}} \pm \frac{2\pi}{\lambda} = \frac{\pi}{a_{zz}}\left(1 \pm \frac{1}{4}\right)$ in Figure S4a. This particular linear combination satisfies the boundary conditions for the (7, ∞) GNR, namely that the wave function must vanish at the first carbon sites *outside* the GNR, which are separated by $(m+1)\frac{a_{zz}}{2} = 4a_{zz}$. Within tight binding, the edge state at the terminus of a (7,∞) GNR is therefore closely linked to the Bloch states of the infinite zigzag edge with crystal wave vectors $k_0^{\pm}$ and its decay constant is given by $\alpha_{k_0} \approx 1.3/\text{nm}$ (with decay constant $2\alpha_{k_0} \approx 2.5/\text{nm}$ for the charge density).

For finite (7,n) GNRs, the edge states of the two termini overlap. This gives rise to a finite energy splitting between a bonding and an anti-bonding linear combination of edge states, as depicted in Figure S4c.

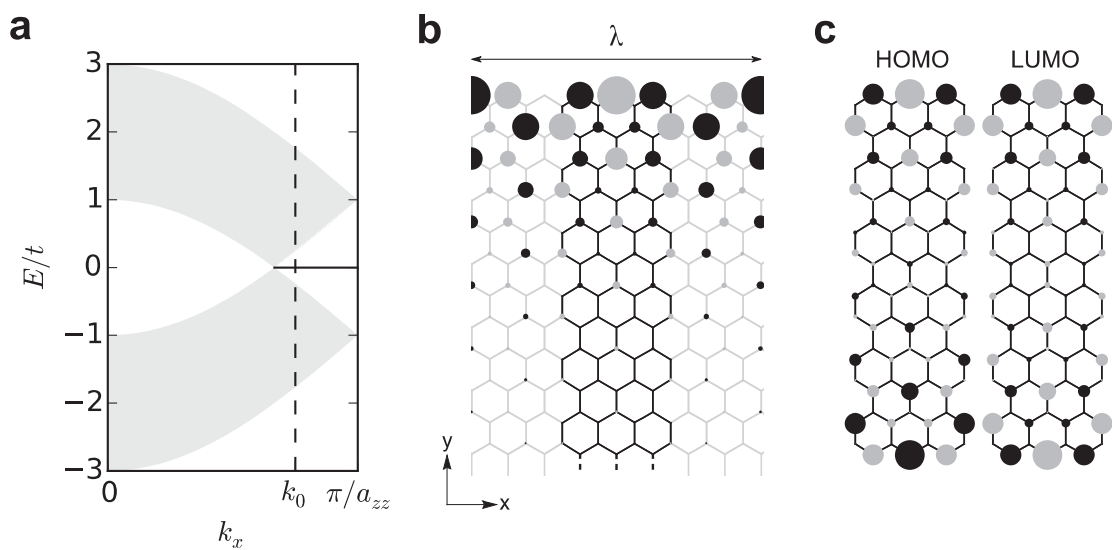

**Figure S4 | Edge states in tight binding. a,** Band structure of graphene projected onto the zigzag direction. The solid black line indicates the states localized at the zigzag edge of a semi-infinite graphene sheet. $k_0 = 0.75\frac{\pi}{a_{zz}}$. **b,** Semi-infinite graphene sheet and linear combination of Bloch waves with wave-length $\lambda = 8a_{zz}$, satisfying the boundary conditions for (7, ∞) GNRs. The circle area is proportional to the electron density, while grey/black indicates the sign of the wave function. **c,** Highest occupied molecular orbital (HOMO, "bonding") and lowest unoccupied molecular orbital (LUMO, "anti-bonding") of (7,12) GNR.